\documentclass[final,5p,times,twocolumn,sort,compress]{elsarticle}

\usepackage[utf8]{inputenc}
\usepackage[T1]{fontenc}
\usepackage{lipsum}
\usepackage{bm}
\usepackage{times}
\usepackage{txfonts}
\usepackage{amssymb,amsbsy,amsmath,amsfonts,amsthm}
\usepackage{graphicx}
\usepackage{float}
\usepackage{color}
\usepackage{morefloats}
\usepackage{rotating}
\usepackage{srcltx}
\usepackage{slashed}
\usepackage{subfigure}
\usepackage{multirow}
\usepackage{verbatim}
\usepackage{hyperref}
\usepackage{tabularx}
\usepackage{adjustbox}
\usepackage{stfloats}
\usepackage[dvipsnames]{xcolor}
\usepackage{nicefrac}
\usepackage{soul}
\usepackage{hyperref}
\usepackage{ulem}
\usepackage[thicklines]{cancel}
\hypersetup{
	colorlinks	=true,
	urlcolor	=blue,
	linkcolor	=blue,
	citecolor	=blue,
	pdftitle	={},
	pdfsubject	={In-medium interactions}
	}

\journal{Physics Letters B}

\begin{document}

\begin{frontmatter}

\title{In-medium $\Lambda N$ interactions with leading order covariant chiral hyperon/nucleon-nucleon forces}

\author[label1]{Ru-You Zheng}

\author[label1]{Zhi-Wei Liu\corref{cor1}}\ead{liuzhw@buaa.edu.cn}

\author[label1,label2,label3,label4,label5]{Li-Sheng Geng\corref{cor1}}\ead{lisheng.geng@buaa.edu.cn}

\author[label6,label7]{Jin-Niu Hu}

\author[label8]{Sibo Wang}

\affiliation[label1]{organization={School of Physics},
            addressline={Beihang University},
            city={Beijing 102206},
            country={China}}
\affiliation[label2]{organization={Sino-French Carbon 
            Neutrality 
            Research Center, \'Ecole Centrale de P\'ekin/School of General Engineering},
            addressline={Beihang University},
            city={Beijing 102206},
            country={China}}
\affiliation[label3]{organization={Peng Huanwu Collaborative 
            Center for Research and Education},
            addressline={Beihang University},
            city={Beijing 102206},
            country={China}}
\affiliation[label4]{organization={Beijing Key Laboratory of 
            Advanced Nuclear Materials and Physics},
            addressline={Beihang University},
            city={Beijing 102206},
            country={China}}
\affiliation[label5]{organization={Southern Center for 
            Nuclear- 
            Science Theory (SCNT)},
            addressline={ Institute of Modern Physics, Chinese Academy of Sciences},
            city={Huizhou 516000},
            country={China}}
\affiliation[label6]{organization={School of Physics},
            addressline={ Nankai University},
            city={Tianjin 300071},
            country={China}}
\affiliation[label7]{organization={Shenzhen Research Institute of Nankai University},
            city={Shenzhen 518083},
            country={China}}
\affiliation[label8]{organization={Department of Physics and Chongqing Key Laboratory for Strongly Coupled 
            Physics},
            addressline={Chongqing University},
            city={Chongqing 401331},
            country={China}}

\cortext[cor1]{Corresponding author.}

\begin{abstract}
In-medium $\Lambda N$ interactions are crucial in hypernuclei and neutron star physics. In this work, we study the in-medium $\Lambda N$ interaction within the relativistic Brueckner-Hartree-Fock (RBHF) framework, employing the leading-order covariant chiral hyperon/nucleon-nucleon forces for the first time. We demonstrate that a consistent description of both the experimental cross-section data and the ‘empirical value' of the $\Lambda$ single-particle potential can be achieved. This contrasts with the majority of studies in the non-relativistic framework, where higher-order two-body chiral forces are typically required. This study offers a new perspective on the in-medium $\Lambda N$ interactions, urgently needed in relativistic \textit{ab initio} hypernuclear physics studies.
\end{abstract}

\end{frontmatter}


\section{Introduction}
\label{introduction}

Hyperon-nucleon $(YN)$ and hyperon-hyperon $(YY)$ interactions are crucial for understanding the role of strange quarks in nuclear physics~\cite{Gal:2016boi,Tolos:2020aln}. They can be classified into free-space interactions, often called bare interactions, and in-medium interactions, which include complex quantum many-body effects. A challenge in nuclear physics has been to elucidate the in-medium properties of hyperons~\cite{Lonardoni:2014bwa,Hiyama:2019kpw,BESIII:2023clq,Crede:2024hur,Achenbach:2024wgy}. The in-medium $YN$ interaction is the most pivotal input for studying hypernuclear physics, hyperon-nucleus reactions, and the equation of state (EoS) of neutron stars. The binding energies of $\Lambda$ hypernuclei suggested that the depth of the $\Lambda$ single-particle potential in the nuclear medium is approximately $27-30$ MeV~\cite{Millener:1988hp,Yamamoto:1988qz,Hashimoto:2006aw}, retaining a relatively large uncertainty. The first observation of hypernuclei-directed flow in high-energy nuclear collisions opens a new venue for studying the $YN$ interaction under finite pressure~\cite{STAR:2022fnj}. The $\Lambda$ single-particle potential serves as a crucial input to elucidate the $\Lambda$ directed flow phenomena in heavy-ion collisions, facilitating a deeper understanding of the underlying dynamics and medium effects~\cite{Nara:2022kbb,Ma:2023nst}. The ‘hyperon puzzle’ associated with neutron stars is a heatedly-discussed topic in nuclear astrophysics. The presence of hyperons in neutron stars seems to be energetically inevitable, however, a naive inclusion of $\Lambda$-hyperons as an additional baryonic degree of freedom softens the EoS such that it fails to support $2 M_{\odot}$ neutron stars~\cite{Lonardoni:2014bwa,Tolos:2020aln,Tong:2024egi}, which have been confirmed by astrophysical observations~\cite{Antoniadis:2013pzd}. This reflects the lack of understanding of the in-medium $\Lambda N$ interactions and highlights the need for further investigations.

There are two main classes of many-body methods for studies of in-medium $YN$ interactions: mean field models and \textit{ab initio} methods. Various mean-field models and phenomenological approaches, such as the Skyrme-Hartree-Fock approach~\cite{Chen:2022ufc,Xue:2023cfm}, the relativistic mean field theory~\cite{Liu:2018img,Rong:2021bim,Guo:2022eqo,Ding:2023qdu,Zhai:2024ggp,Ding:2024gdv}, the relativistic Hartree-Fock theory~\cite{Long:2011hx, Ding:2022gbu}, the quark mean-field model~\cite{Hu:2013cqa,Xing:2017zaj,Hu:2021ket}, and the optical potential model~\cite{Friedman:2022bpw,Friedman:2023ucs} have been employed to obtain the in-medium $YN$ interactions. In these models, the in-medium $YN$ interactions are determined by fitting the calculated $\Lambda$ separation energies to the experimental values of known hypernuclei or turning to models such as the naive quark model. On the other hand, in \textit{ab initio} calculations, one uses bare interactions to construct the in-medium effective interactions~\cite{Machleidt:2023jws}. The \textit{ab initio} calculations do not contain any free parameters and are theoretically more sound and predictive. Recently, the non-relativistic Brueckner-Hartree-Fock theory (BHF) has been widely employed to construct the $YN$ effective interactions in hyperon matter~\cite{Reuber:1993ip,Schulze:1998jf,Vidana:1999jm,Kohno:1999nz,Fujiwara:1999fe,Yamamoto:2000jh,
Kohno:2009sc,Schulze:2011zza,Kohno:2013vxa,Haidenbauer:2014uua,Petschauer:2015nea,Haidenbauer:2016moh,Haidenbauer:2016vfq,Haidenbauer:2018gvg,Kohno:2018gby,Haidenbauer:2019boi,Gerstung:2020ktv,Haidenbauer:2023qhf}. 
The widely used $YN$ bare interactions include the Nijmegen and Jülich meson exchange potentials~\cite{Maessen:1989sx,Rijken:1998yy,Holzenkamp:1989tq,Reuber:1993ip,Haidenbauer:2005zh,Rijken:2006en,Rijken:2010zzb}, the $SU(3)$ or $SU(6)$ quark model potentials~\cite{Kohno:2009sc}, and the non-relativistic chiral $YN$ interactions~\cite{Haidenbauer:2013oca,Haidenbauer:2019boi,Haidenbauer:2023qhf}. In the studies of neutron stars within the BHF framework, repulsive three-body forces are needed such that one can obtain massive neutron stars consistent with astrophysical observations~\cite{Yamamoto:2014jga,Haidenbauer:2016vfq,Logoteta:2019utx,Kochankovski:2022hjl,Vidana:2024ngv}.
 
On the other hand, the relativistic Brueckner-Hartree-Fock (RBHF) method has seen substantial progress in nuclear matter studies~\cite{Brockmann:1990cn,Tong:2018qwx,Wang:2020exc,Wang:2020dov,Wang:2021mvg,Wang:2022sev,Tong:2022yml,Tong:2022tlt,Wang:2022sev,Wang:2022fqt,Wang:2022xlr,Qu:2023cpm,Wang:2023owh,Wang:2023zdc,Wang:2024jtl,Qin:2024zzi}. In Ref~\cite{Brockmann:1990cn}, with two-body interactions only, the RBHF approach could yield saturation properties closer to the empirical values, contrasting with the BHF theory. In particular, the RBHF theory based on the leading-order (LO) covariant chiral nucleon-nucleon ($NN$) interaction can already describe the saturation properties of symmetric nuclear matter (SNM)~\cite{Zou:2023quo}. These achievements underscore the importance of \textit{ab initio} calculations in the RBHF framework employing microscopic $YN$ interactions. However, at present, the bare $YN$ interaction used in the RBHF is based on the phenomenological Jülich potential~\cite{Sammarruca:2009wn,Hu:2014wja,Katayama:2015dga}, which suffers from limitations such as the absence of a microscopic foundation and the inability to estimate associated theoretical uncertainties. From this perspective, developing a microscopic relativistic $YN$ force within the RBHF theory represents a crucial step forward.  

Recently, we have successfully constructed the LO relativistic chiral $YN$ and $YY$ interactions based on the covariant chiral effective field theory (ChEFT)~\cite{Li:2016mln,Li:2018tbt,Liu:2020uxi,Song:2021yab,Liu:2022nec}. In contrast to the non-relativistic chiral $YN$ interactions that were more commonly used in earlier studies~\cite{Haidenbauer:2013oca,Haidenbauer:2019boi,Haidenbauer:2023qhf}, the covariant treatment of baryon spinors in this approach not only preserves all relevant symmetries but also exhibits faster convergence, thereby providing the much-needed microscopic bare $YN$ interaction inputs for the RBHF studies. It is worth noting that, unlike in Ref.~\cite{Li:2016mln,Liu:2020uxi,Liu:2022nec} where $M_B$ is treated as the average baryon mass, studying the in-medium $\Lambda N$ interaction requires distinguishing between the masses of $\Lambda$, $\Sigma$, and $N$. Therefore, we will re-derive the covariant chiral $YN$ interactions incorporating physical baryon masses in this work.

In this study, we first re-derive the covariant chiral $YN$ interactions with physical baryon masses and further investigate the in-medium properties of this interaction within the RBHF theory. This paper is organized as follows. Sec.~\ref{Theoretical Framework} briefly introduces the RBHF theory based on the LO covariant chiral $YN$ force. We analyze the results in Sec.~\ref{Results and Discussions}, followed by a summary and outlook in Sec.~\ref{Summary and outlook}.

\section{Theoretical Framework}\label{Theoretical Framework}

\subsection{Hyperon matter in the relativistic Brueckner-Hartree-Fock theory}

We study the properties of hyperon matter employing the RBHF theory. When hyperon degrees of freedom are considered, the in-medium Dirac equation can be expressed as follows:
\begin{equation}\label{Dirac equation}
(\boldsymbol{\alpha}\cdot\boldsymbol{p}+\beta M_B+\beta\mathcal{U})u(\boldsymbol{p},\lambda)=E_B(\boldsymbol{p})u(\boldsymbol{p},\lambda),
\end{equation}
where $\boldsymbol{\alpha}$ and $\beta$ are the Dirac matrices, $u(\boldsymbol{p}, \lambda)$ is the Dirac spinor with
momentum $\boldsymbol{p}$ and helicity $\lambda$. $M_B$ is the physical mass of the free baryon and $E_B(\boldsymbol{p})$ is the single-particle energy. The single-particle potential operator, $\mathcal{U}$, encapsulates the primary medium effects. Due to its weakness compared to the scalar field and the timelike component of the vector fields, the spacelike component of the vector fields can be neglected. As a result, we use the following ansatz for the single-particle potential~\cite{Brockmann:1990cn}
\begin{equation}
\mathcal{U}=U_S^B+\gamma^0 U_0^B,
\end{equation}
where $U_S^{B}$ and $U_0^{B}$ denote the scalar and the timelike component of the vector potentials~\footnote{In this work, $U_S^{N}$ and $U_0^{N}$ are taken from the RBHF calculation with the covariant chiral $NN$ interaction in Ref.~\cite{Zou:2023quo}.}, respectively. It has been demonstrated that the momentum dependence of these potentials is sufficiently weak to be neglected within the RBHF framework~\cite{Wang:2021mvg}. In this work, the continuous choice for the intermediate-state spectrum is adopted.

The solution of the in-medium Dirac equation reads
\begin{equation}
u(\boldsymbol{p}, \lambda)=\sqrt{\frac{E_B^*(\boldsymbol{p})+M_B^*}{2 E_B^*(\boldsymbol{p})}}\binom{1}{\frac{2 \lambda p}{E_B^*(\boldsymbol{p})+M_B^*}} \chi_\lambda,
\end{equation}
where $\chi_\lambda$ is the Pauli spinor helicity basis. The covariant normalization is $u^{\dagger}(\boldsymbol{p}, \lambda) u(\boldsymbol{p}, \lambda)=1$. $M_B^*$ and $E_B^*$ represent the baryon effective mass and energy, respectively, which are defined as \begin{equation}
M_B^*=M_B+U_S^B, \quad E_B^*(\boldsymbol{p})=E_B(\boldsymbol{p})-U_0^B,
\end{equation} the single-particle energy $E_B(\boldsymbol{p})=\sqrt{M_B^{* 2}+\boldsymbol{p}^{ 2}}+U_0^B$.
Once the in-medium Dirac spinor is obtained, the in-medium baryon-baryon ($BB$) interaction matrix elements can be calculated. Detailed calculations will be presented in Sec.~\ref{LO YN}.

In the RBHF framework, the $G$-matrix is obtained by solving the in-medium relativistic scattering equation, i.e., the Bethe-Goldstone equation,
\begin{equation}\label{BS equation}
\begin{aligned}
 &\langle B_3B_4| G(\zeta)|B_1B_2 \rangle=\langle B_3B_4| V|B_1B_2 \rangle \\
 &+\sum_{B_5B_6}\langle B_3B_4| V|B_5B_6\rangle \frac{Q_{B_5B_6}}{\zeta-E_{B_5 B_6}}\langle B_5B_6| G(\zeta)|B_1B_2 \rangle,
\end{aligned}
\end{equation}
with $B_1B_2,B_3B_4,B_5B_6=\Lambda N,\Sigma N$ in the strangeness $S=-1$ system. Here, $Q_{B_5B_6}$ denotes the Pauli projection operator which excludes the $B_5B_6$ intermediate state below their respective Fermi momentum, $k_F^{\left(B_5\right)}$ and $k_F^{\left(B_6\right)}$. $E_{B_5B_6}$ denotes the energy of the intermediate state. 
The starting energy $\zeta$ for an initial $\Lambda N$-state with momenta $\boldsymbol{p}_\Lambda$ and $\boldsymbol{p}_N$ is given by $\zeta=E_{\Lambda}\left(\boldsymbol{p_{\Lambda}}\right)+E_N\left(\boldsymbol{p_N}\right)$. 

The $\Lambda$ single-particle potential $U_\Lambda(p_\Lambda)$, which is calculated in terms
of the $\Lambda N$ $G$ matrix formally in the usual way~\cite{Reuber:1993ip,Haidenbauer:2018gvg}:
\begin{equation}\label{Lambda spe}
U_{\Lambda}\left(p_\Lambda\right)=\int_{p_N\leqslant k^N_{\mathrm{F}}}\mathrm{~d}^3 p_N~{\rm Tr}~\left\langle\boldsymbol{p_\Lambda},\boldsymbol{p_N}\right|G_{\Lambda N}\left(\zeta\right)\left|\boldsymbol{p}_{\Lambda},\boldsymbol{p}_N\right\rangle,
\end{equation}
where Tr denotes the trace in spin- and isospin-space. In the no-sea approximation, the above integral is only performed for the single-nucleon states in the Fermi sea, $\left|\boldsymbol{p}_N\right|\leqslant k_F^{N}$. To facilitate a partial wave decomposition of Eq.~(\ref{BS equation}), the Pauli projection operator $Q_{B_5B_6}$ and the total momentum of the $YN$ states, $\boldsymbol{P}=\boldsymbol{p}_\Lambda+\boldsymbol{p}_N$, have to be approximated by their angle-averages $\bar{Q}_{B_5B_6}$ and $\overline{P^2}$ (The detailed expressions can be found in Refs.~\cite{Reuber:1993ip,Vidana:1999jm,Hu:2014wja}.).

On the other hand, the single-particle potential in the RBHF theory can be expressed as
\begin{equation}\label{lambda defined}
U_{\Lambda}\left(p_{\Lambda}\right)=\bar{u}(\boldsymbol{p}_\Lambda, 1 / 2) \mathcal{U} u(\boldsymbol{p}_\Lambda, 1 / 2)=\frac{M_{\Lambda}^*}{\sqrt{\left(M_{\Lambda}^*\right)^2+p_{\Lambda}^2}} U_S^{\Lambda}+U_0^{\Lambda}.
\end{equation}

Eqs.~(\ref{Dirac equation}), (\ref{BS equation}), and (\ref{lambda defined}) constitute a set of coupled equations that need to be solved self-consistently. Starting from certain initial values of $U_S^{\Lambda(0)}$ and $U_0^{\Lambda(0)}$, one solves the in-medium Dirac equation (\ref{Dirac equation}) to obtain the Dirac spinors. Next, one solves the in-medium scattering equation (\ref{BS equation}) to obtain the $G_{\Lambda N}$ matrix and uses the integrals in Eq.(\ref{Lambda spe}) to obtain $U_{\Lambda}\left(p_{\Lambda}\right)$. With Eq.(\ref{lambda defined}), one then has a new set of fields, $U_S^{\Lambda(1)}$ and $U_0^{\Lambda(1)}$, to be used in the next iteration. Once $U_S^{\Lambda}$ and $U_0^{\Lambda}$ of the single-particle potential converge, the effective $\Lambda N$ interaction $U_{\Lambda}\left(p_{\Lambda}=0\right)$ can be calculated. This study focuses on a small fraction of $\Lambda$ baryons in hyperon matter, where a mixture of nucleons and $\Lambda$ hyperons can be considered `metastable'~\cite{Sammarruca:2009wn}. When solving the in-medium $G_{\Lambda N}$ matrix, as described in Eq. (\ref{BS equation}), a coupled-channel formalism for $YN$ interactions ($S=-1$) is employed, where intermediate states involving the $\Sigma N$ channel. However, the $\Sigma$ hyperon density is assumed to be zero; thus, the in-medium effects on the $\Sigma$ baryon are not considered. 

\subsection{Hyperon-nucleon interactions in the leading-order covariant ChEFT}\label{LO YN}
The bare $YN$ interactions are one of the most important
inputs in studies of hyperon matter. Here, we briefly introduce the covariant ChEFT for the $YN$ interaction. The LO $YN$ interactions include non-derivative four baryon contact terms (CT) and one-pseudoscalar-meson exchange (OPME) potentials~\cite{Li:2016mln}, i.e., $V_{\mathrm{LO}}=V_{\mathrm{CT}}+V_{\mathrm{OPME}}$. The CT potential reads
\begin{equation}
V_{\mathrm{CT}}^{\mathrm{YY}^{\prime}}=C_i^{\mathrm{YY}{ }^{\prime}}\left(\bar{u}_3 \Gamma_i u_1\right)\left(\bar{u}_4 \Gamma_i u_2\right),
\end{equation}
where $C_i^{\mathrm{YY}{ }^{\prime}}$ are the LECs. The superscript $YY^{\prime}$ denotes the hyperons in the reaction of $YN$ $\rightarrow$ $Y^{\prime}N$. The Clifford algebra is $\Gamma_1=1,\Gamma_2=\gamma^\mu,\Gamma_3=\sigma^{\mu v},\Gamma_4=\gamma^\mu \gamma_5$, and $\Gamma_5=\gamma_5$. The CT potentials are first calculated in the helicity basis and then projected into different partial waves in the $|LSJ\rangle$ basis, which read
\begin{subequations}
 \begin{align}  V_{^1S_0}=\xi&\left[C_{1S0}^{\mathrm{YY^\prime}}\left(1+R_1^{\prime} R_1R_2^{\prime} R_2\right)+\hat{C}_{1S0}^{\mathrm{YY^\prime}} \left(R_1^{\prime}R_2^{\prime}+R_1R_2\right)\right],\\
  V_{^3S_1}=\xi&\left[\frac{1}{9}C_{3S1}^{\mathrm{YY^\prime}}\left(9+R_1^{\prime}R_1R_2^{\prime}R_2\right)+\frac{1}{3}\hat{C}_{3S1}^{\mathrm{YY}^{\prime}}\left(R_1^{\prime}R_2^{\prime}+R_1R_2\right)\right],\\
  V_{^3D_1}=\xi&\left[\frac{8}{9}C_{3S1}^{\mathrm{YY}^{\prime}}  R_1^{\prime}R_1R_2^{\prime}R_2\right],\\
  V_{^3SD_1}=\xi&\left[\frac{2\sqrt{2}}{9}C_{3S1}^{\mathrm{YY}^{\prime}} R_1^{\prime}R_1R_2^{\prime}R_2+\frac{2\sqrt{2}}{3}\hat{C}_{3S1}^{\mathrm{YY}^{\prime}}R_1R_2\right],\\
  V_{^3DS_1}=\xi&\left[\frac{2\sqrt{2}}{9}C_{3S1}^{\mathrm{YY}^{\prime}} R_1^{\prime}R_1R_2^{\prime}R_2+\frac{2\sqrt{2}}{3}\hat{C}_{3S1}^{\mathrm{YY}^{\prime}}R_1^{\prime}R_2^{\prime}\right],\\
  V_{^3P_0}=-\xi&\left[C_{3P0}^{\mathrm{YY}^{\prime}}\left(R_1^{\prime} R_1+R_2^{\prime}R_2+R_1^{\prime}R_2+R_2^{\prime}R_1\right)\right],\\
  V_{^3P_1}=-\frac{2}{3}&\xi\left[\frac{\left(C_{1S0}^{\mathrm{YY}^{\prime}}-\hat{C}_{1S0}^{\mathrm{YY}^{\prime}}\right)}{2}\left(R_1^{\prime}R_1+R_2^{\prime}R_2\right)\right.\nonumber\\
  &\left.+\frac{\left(C_{1S0}^{\mathrm{YY}^{\prime}}-\hat{C}_{1S0}^{\mathrm{YY}^{\prime}}\right)}{2} \left(R_1^{\prime}R_2+R_2^{\prime}R_1\right)\right],\\
  V_{^1P_1}= -\frac{1}{3}&\xi\left[\frac{\left(C_{3S1}^{\mathrm{YY}^{\prime}}-\hat{C}_{3S1}^{\mathrm{YY}^{\prime}}\right)}{2}\left(R_1^{\prime}R_1+R_2^{\prime}R_2\right)\right.\nonumber\\
  &\left.+\frac{\left(C_{3S1}^{\mathrm{YY}^{\prime}}-\hat{C}_{3S1}^{\mathrm{YY}^{\prime}}\right)}{2} \left(R_1^{\prime}R_2+R_2^{\prime}R_1\right)\right],
\end{align}
\end{subequations} where $\xi=4 \pi \sqrt{\frac{(E^{Y^\prime}_{p^\prime}+M_Y)(E^Y_p+M_Y)(E^N_{p^\prime}+M_N)(E^N_p+M_N)}{{16 E^{Y^\prime}_{p^\prime} E^Y_p E^N_{p^\prime} E^N_p}}}$, $R_1=p/(E^Y_p+M_Y), R_2=p/(E^N_p+M_N), R_1^{\prime}=p^{\prime}/(E^{Y^\prime}_{p^\prime}+M_Y)$, and $R_2^{\prime}=p^{\prime}/(E^N_{p^\prime}+M_N)$. It should be noticed that the quantities $\xi$, $R_1$, $R_2$, $R^\prime_1$, and $R^\prime_2$ require medium-induced modifications through the substitutions $M\rightarrow M^*$ and $E\rightarrow E^*$ when going from hyperon-nucleon scattering in vacuum to the in-medium calculation.

The OPME potentials have the following form,
\begin{equation}
V_{\mathrm{OPME}}=N_{B_1 B_3 \phi} N_{B_2 B_4 \phi} \frac{\left(\bar{u}_3 \gamma^\mu \gamma_5 q_\mu u_1\right)\left(\bar{u}_4 \gamma^v \gamma_5 q_v u_2\right)}{q^2-m^2} \mathcal{I}_{B_1 B_2 \rightarrow B_3 B_4},
\end{equation}
where $q=\left(E_{p^{\prime}}-E_p, \boldsymbol{p^{\prime}}-\boldsymbol{p}\right)$ represents the four-momentum transferred, and $m$ is the mass of the exchanged pseudoscalar
meson. $N_{B_1 B_3 \phi}N_{B_2 B_4 \phi}$ and $\mathcal{I}_{B_1 B_2 \rightarrow B_3 B_4}$ are the $SU(3)$ coefficients and isospin factors, respectively, listed in Refs.~\cite{Polinder:2006zh,Li:2016mln}. We can obtain $V_\mathrm{OPME}$ in the $|L S J\rangle$ basis following the same procedure as the CT potential.

In addition, to avoid ultraviolet divergence in numerical evaluations, the $YN$ potentials are regularized with an exponential form factor,
\begin{equation}
f_{\Lambda_F}\left(\boldsymbol{p}, \boldsymbol{p}^{\prime}\right)=\exp \left[-\left(\frac{\boldsymbol{p}}{\Lambda_F}\right)^{2 n}-\left(\frac{\boldsymbol{p}^{\prime}}{\Lambda_F}\right)^{2 n}\right],
\end{equation}
where $n = 2$ and $\Lambda_F$ is the cutoff momentum. We consider cutoff values from 550 to 700 MeV in the present work. The resulting uncertainties indicate the cutoff dependence and thus provide a lower bound on theoretical uncertainties.

\section{Results and Discussions}\label{Results and Discussions}

As explained in the previous section, distinguishing the physical masses of $\Lambda$, $\Sigma$, and $N$ is crucial for studying the in-medium interaction in the RBHF framework. To address this, we have re-derived the covariant chiral $YN$ interaction, explicitly considering physical baryon masses $M_N=938.918$ MeV, $M_\Lambda=1115.68$ MeV, and $M_\Sigma=1193.1$ MeV. In the contact terms, there are 12 independent LECs, namely, $C_{1 S 0}^{\Lambda \Lambda}$, $C_{1 S 0}^{\Sigma \Sigma}$, $C_{3 S 1}^{\Lambda \Lambda}$, $C_{3 S 1}^{\Sigma \Sigma}$, $C_{3 S 1}^{\Lambda \Sigma}$, $\hat{C}_{1 S 0}^{\Lambda \Lambda}$, $\hat{C}_{1 S 0}^{\Sigma \Sigma}$, $\hat{C}_{3 S 1}^{\Lambda \Lambda}$, $\hat{C}_{3 S 1}^{\Sigma \Sigma}$, $\hat{C}_{3 S 1}^{\Lambda \Sigma}$, $C_{3 P 0}^{\Lambda \Lambda}$, and $C_{3 P 0}^{\Sigma \Sigma}$. These 12 LECs were determined by fitting the 36 $YN$ scattering data~\cite{Engelmann:1966npz,Sechi-Zorn:1968mao,Alexander:1968acu,Eisele:1971mk, Kadyk:1971tc,Hauptman:1977hr}. The corresponding values of the LECs and $\chi^2$ are listed in Table~\ref{lecs}. The obtained $\chi^2$ is comparable with the sophisticated phenomenological models and the non-relativistic next-to-leading order (NLO) ChEFT. We employ the covariant chiral $YN$ force, as listed in Table~\ref{lecs}, to calculate cross sections and compare them with those obtained using average masses~\cite{Liu:2020uxi}. Subsequently, we calculate the in-medium $\Lambda N$ interactions. It has been demonstrated that the simultaneous description of $NN$ and $YN$ scattering data with $SU(3)$ symmetric LECs is impossible at LO ChEFT~\cite{Petschauer:2015nea,Li:2016mln,Liu:2020uxi}. Therefore, we must use different sets of LECs. To be consistent with the LO $YN$ interaction, we employ the LO $NN$ interaction with the same cutoff from Ref.~\cite{Zou:2023quo}.

\begin{table*}[bp]
\centering
\setlength{\tabcolsep}{4.8pt}
\caption{Low-energy constants (in units of $10^4$ GeV$^{-2}$) and $\chi^2$ obtained for two cutoffs of $\Lambda_F=550$ and 700 MeV in the covariant ChEFT. These LECs are determined by fitting the $S=-1$ $YN$ scattering data.}
\begin{tabular}{lccccccccccccc}
\hline \hline$\Lambda_F$ & $C_{1 S 0}^{\Lambda \Lambda}$ & $C_{1 S 0}^{\Sigma \Sigma}$ & $C_{3 S 1}^{\Lambda \Lambda}$ & $C_{3 S 1}^{\Sigma \Sigma}$ & $C_{3 S 1}^{\Lambda \Sigma}$ & $\hat{C}_{1 S 0}^{\Lambda \Lambda}$ & $\hat{C}_{1 S 0}^{\Sigma \Sigma}$ & $\hat{C}_{3 S 1}^{\Lambda \Lambda}$ & $\hat{C}_{3 S 1}^{\Sigma \Sigma}$ & $\hat{C}_{3 S 1}^{\Lambda \Sigma}$ & $C_{3 P 0}^{\Lambda \Lambda}$ & $C_{3 P 0}^{\Sigma \Sigma}$ & $\chi^2$\\
\hline 550 & -0.0715 & -0.1001 & 0.0322 & 0.1228 & 0.0555 & 1.7864 & 2.1145 & 0.7677 & -1.6686 & 1.2780 & -0.8742 & -1.3728 & 15.85\\
700 & -0.0222 & -0.0397 & 0.0420 & 0.1138 & 0.0195 & 2.2580 & 2.6327 & -0.1415 & -0.0438 & 0.6171 & -1.5247 & -2.0052 & 15.74 \\
\hline \hline
\end{tabular}
\label{lecs}
\end{table*}

\begin{figure*}[bp]
    \centering
    \includegraphics[width=1.0\linewidth]{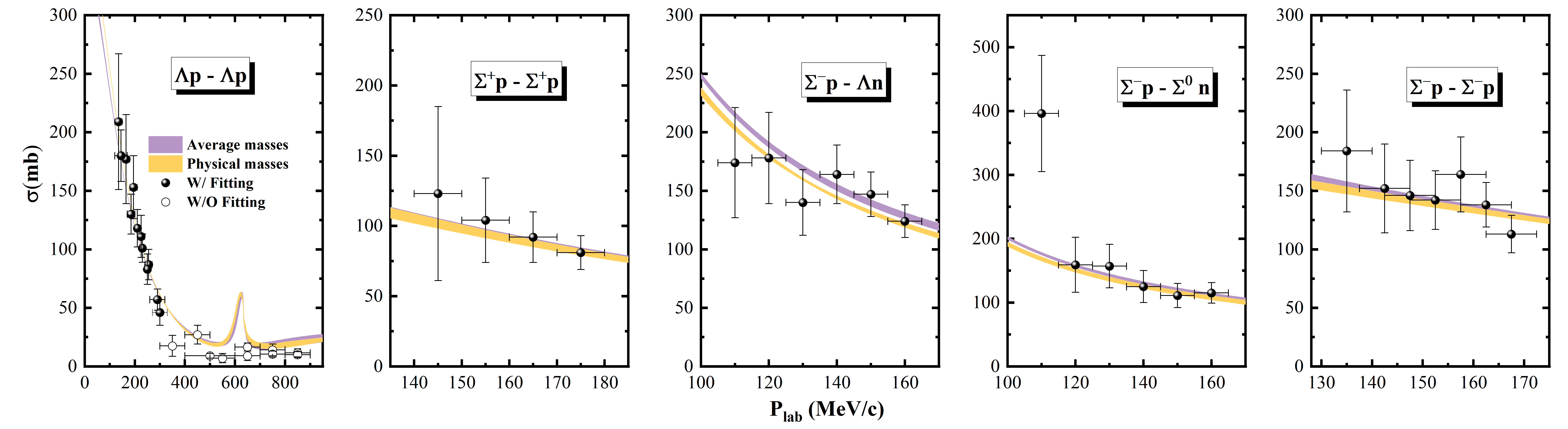}
    \caption{Cross sections obtained with the LO covariant chiral $YN$ interaction as functions of the laboratory momentum for $\Lambda_F=550-700$ MeV. The yellow bands represent the results obtained with physical baryon masses, and the purple bands represent the results obtained with the average baryon masses~\cite{Liu:2020uxi}. The experimental cross sections are taken from Refs.~\cite{Engelmann:1966npz,Sechi-Zorn:1968mao,Alexander:1968acu,Eisele:1971mk,Kadyk:1971tc,Hauptman:1977hr}.}
    \label{cross section}
\end{figure*}

The cross sections obtained with physical and average masses are presented in Fig.~\ref{cross section}. The bands represent the variation of the cross sections with the cutoff in the region of $\Lambda_F=550-700$ MeV. One can reproduce the experimental data using the covariant chiral $YN$ interaction constructed with the physical baryon masses. The cross sections obtained with physical masses differ slightly from those obtained with average masses. Our results for the $\Sigma^{-} p \rightarrow \Lambda n$, $\Sigma^{-} p \rightarrow \Sigma^{0} n$, and $\Sigma^{-} p \rightarrow \Sigma^{-} p$ reactions show an overall downward shift compared to the case of the average masses. The cross sections for $\Lambda p\rightarrow \Lambda p$ agree with the data even up to the $\Sigma N$ threshold. For $P_\mathrm{lab}$ below 200 MeV/c, the agreement with the experimental central values is improved when using the physical masses compared to the average masses. For the $\Sigma^{-} p \rightarrow \Lambda n$, the physical masses results agree better with the experimental data for $P_\mathrm{lab}$ below 130 MeV/c.

Table~\ref{T.ulp=0} summarizes the $\Lambda$ single-particle potential $U_\Lambda(p_\Lambda = 0)$ at the saturation density of SNM ($k_F=1.35$ fm$^{-1}$), calculated with our LO covariant chiral $YN$ and $NN$ interactions for the cutoff range studied. The results obtained from the various non-relativistic (NonRel.) chiral $YN$ forces~\cite{Haidenbauer:2014uua,Haidenbauer:2019boi,Mihaylov:2023ahn,Haidenbauer:2023qhf}, the Jülich potentials from 2005 (Jul05)~\cite{Haidenbauer:2005zh} and 1994 (Jul94)~\cite{Reuber:1993ip}, and the Nijmegen NSC97f model~\cite{Rijken:1998yy} are also presented for comparison. Our LO(700) result, along with the non-relativistic chiral NLO13(500), NLO19(650) results, and the Jul94 result are consistent with the empirical value of $-$27 to $-$30 MeV. Note that in the non-relativistic chiral NLO13 and NLO19 calculations, the $NN$ interactions used were phenomenological~\cite{Haidenbauer:2014uua,Haidenbauer:2019boi}.
~In our results, the in-medium effective interactions in the ${ }^1 S_0$ and ${ }^3 S_1+{ }^3 D_1$ partial waves are mainly attractive, while ${ }^3 P_0$, ${ }^1 P_1$, and ${ }^3 P_1$ mainly repulsive. This behavior is consistent with the NLO13, Jul94, and NSC97f results. As the cutoff increases, the attractive contribution in the $^1{S}_{0}$ and ${ }^3 S_1+{ }^3 D_1$ partial waves gradually increases, while the repulsive contribution of the $P$-wave weakens. The repulsive contributions of ${ }^1 P_1$ and ${ }^3 P_1$ are larger than those of ${ }^3 P_0$.
\begin{table*}[bp]
\centering
\setlength{\tabcolsep}{16.8pt}
\caption{The partial-wave contributions to the $\Lambda$ single-particle potential (in units of MeV) in different $YN$ models at the nuclear saturation density. The latest non-relativistic chiral result is denoted as `NLO19$^*$(500-650)', where the non-relativistic chiral $YN$ interaction is constrained by the combined analysis of low-energy scattering data and femtoscopic correlation data~\cite{Mihaylov:2023ahn}, and the uncertainty is determined by adding in quadrature the resulting ones from the systematic and statistical uncertainties.}
\label{T.ulp=0}
\begin{tabular}{lccccccc}
\hline \hline $YN$ potential(cutoff)& ${ }^1 S_0$ & ${ }^3 S_1+{ }^3 D_1$ & ${ }^3 P_0$ & ${ }^1 P_1$ & ${ }^3 P_1$ & Total \\
\hline  This work LO(550) & $-$18.8& $-$23.5 &1.4 &4.3 & 11.4&$-$25.1 \\
        This work LO(700) &$-$24.0 &$-$12.2 & 0.6& 1.0 &6.3 & $-$29.3\\
        NonRel.$-$ LO(550)~\cite{Haidenbauer:2014uua} & $-$12.5 & $-$26.6&$-$1.6 &1.5 &1.8 &$-$38.0 \\ 
        NonRel.$-$ LO(700)~\cite{Haidenbauer:2014uua} &$-$11.6 & $-$23.1&$-$1.9 &1.5 &1.6 & $-$34.4\\
        NonRel.$-$ NLO13(500)~\cite{Haidenbauer:2014uua} & $-$15.3&$-$15.8 & 1.1& 2.3&1.1 & $-$28.2\\
        NonRel.$-$ NLO13(650)~\cite{Haidenbauer:2014uua} &$-$11.6 & $-$13.4&0.8 & 1.8&0.7 & $-$23.2\\
        NonRel.$-$ NLO19(500)~\cite{Haidenbauer:2019boi} & $-$12.5&$-$28.0 & $-$&$-$ &$-$ & $-$39.3\\
        NonRel.$-$ NLO19(650)~\cite{Haidenbauer:2019boi} &$-$11.1 & $-$19.7&$-$ &$ -$&$-$ & $-$29.2\\
        NonRel.$-$ NLO19$^*$(500-650)~\cite{Mihaylov:2023ahn} &$-$ & $-$&$-$ &$ -$&$-$ & $-36.3_{-6.3}^{+2.8}$\\
        NonRel.$-$ $\mathrm{N}^{2}$LO(500)~\cite{Haidenbauer:2023qhf} &$-$ & $-$&$-$ &$ -$&$-$ & $-$33.1\\
        NonRel.$-$ $\mathrm{N}^{2}$LO(600)~\cite{Haidenbauer:2023qhf} &$-$ & $-$&$-$ &$ -$&$-$ & $-$37.8\\
        Jul94 &$-$3.0 &$-$29.1 &0.89 & 1.7&4.1 & $-$27.3\\
        Jul05 & $-$11.7& $-$37.0& $-$0.9& 0.2&3.2 &$-$50.6 \\
        NSC97f~\cite{Yamamoto:2000jh}&$-$14.6 &$-$23.1 & 0.5& 2.4&4.6 &$-$32.4 \\
\hline \hline
\end{tabular}
\end{table*}

\begin{figure}[htbp]
\centering
\includegraphics[width=0.9\linewidth]{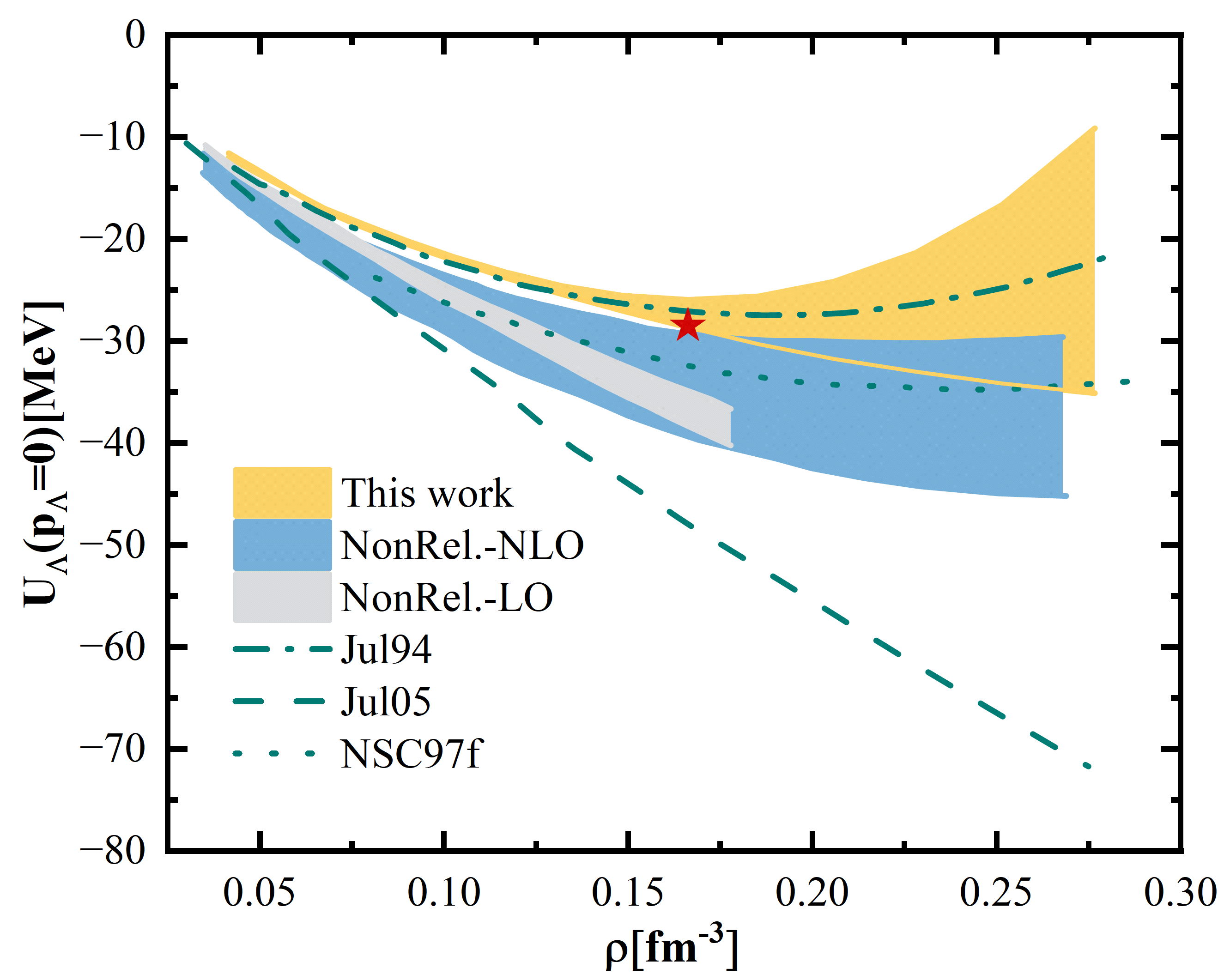}
\caption{The $\Lambda$ single-particle potential $U_\Lambda(p_\Lambda = 0)$ as a function of density in SNM. The yellow band represents the results obtained with the LO covariant chiral $YN$ interaction for cutoffs of $\Lambda_F$ = 550 (upper boundary) and 700 MeV (lower boundary), respectively. The blue band shows the NLO non-relativistic chiral $YN$ interaction results for cutoffs of $\Lambda_F$ = 500 to 650 MeV, while the gray band is the LO non-relativistic chiral $YN$ interaction results for $\Lambda_F$ = 550 to 700 MeV~\cite{Haidenbauer:2014uua,Haidenbauer:2019boi}. The dash-dotted curve is the result of the meson-exchange model Jul94, and the dashed curve is the result of the Jul05 potential~\cite{Hu:2014wja}. The dotted curve is the result of the Nijmegen NSC97f potential~\cite{Rijken:1998yy}, taken from Ref.~\cite{Yamamoto:2000jh}. The red star indicates the `empirical value'.}
\label{ulp=0}
\end{figure}

In Fig.~\ref{ulp=0}, we show the density dependence of the $\Lambda$ single-particle potential $U_\Lambda(p_\Lambda = 0)$ for SNM. The results are calculated with the LO covariant chiral $NN$ and $YN$ potentials in the RBHF framework, compared with the non-relativistic LO, NLO, Jul94, Jul05, and NSC97f results. The covariant chiral result indicates an onset of repulsion already at moderate densities, similar to the $NN$ EoS with the LO covariant chiral $NN$ force~\cite{Zou:2023quo}, but dependent on the cutoff. It is seen that $U_\Lambda(p_\Lambda = 0)$ becomes more attractive as the cutoff $\Lambda_F$ increases (the lower and upper boundaries of the yellow band represent the results for cutoffs of $\Lambda_F=700$ and 550 MeV, respectively). At lower densities, we observe a weak cutoff dependence. In contrast, at higher densities, $U_\Lambda(p_\Lambda = 0)$ strongly depends on the momentum cutoff $\Lambda_F$. The origin of such an exacerbation of cutoff dependence remains unclear at present. Within uncertainties, our results are consistent with the Jul94 results. However, the non-relativistic LO result becomes more attractive as the density increases. The comparison with the results from meson-exchange NSC97f and non-relativistic LO shows that our $\Lambda$ single-particle potential exhibits stronger repulsion. This implies that $\Lambda$ hyperons would emerge at higher densities -- a feature favoring massive neutron star formation. The density dependence predicted by the non-relativistic NLO chiral force is similar to that of the NSC97f potential. In addition, the Jul05 result exhibits a significant density dependence, which can be traced to the strong tensor component of the $\Lambda N-\Sigma N$ interaction~\cite{Hu:2014wja,Haidenbauer:2014uua} (see Table~\ref{T.ulp=0}).

\begin{figure}[htbp]
    \centering
    \includegraphics[width=0.9\linewidth]{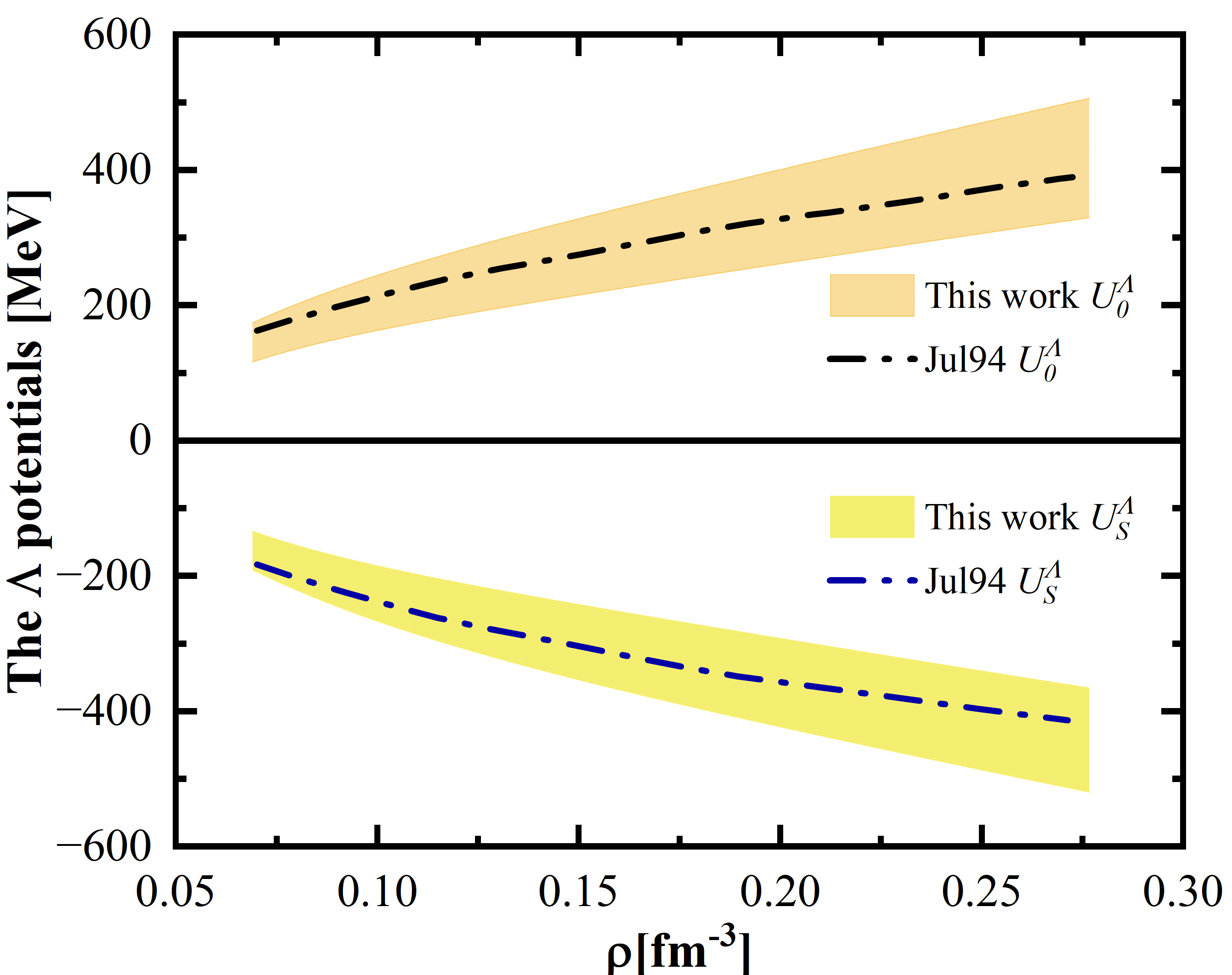}
    \caption{Scalar and vector components of the $U_\Lambda(p_\Lambda = 0)$ obtained in the RBHF theory with the covariant chiral $YN$ and Jul94 potentials. The bands represent the variation of the results with the cutoff, where the lower and upper boundaries of $U_S^\Lambda$ ($U_0^\Lambda$) represent the results for the cutoffs of $\Lambda_F=550$ and 700 MeV (700 and 550 MeV), respectively.}
    \label{usu0}
\end{figure}

The scalar and vector components of the $U_\Lambda(p_\Lambda = 0)$ as functions of the density are shown in Fig.~\ref{usu0}. As the density increases, the scalar potential $U_{\mathrm{S}}^{\Lambda}$ becomes more attractive, while the vector potential $U_{\mathrm{0}}^{\Lambda}$ becomes more repulsive. Similar to the previous results, the scalar and vector potentials obtained with the covariant chiral $YN$ interaction agree with those of the Jul94 potential within uncertainties. However, the $U_{\mathrm{S}}^{\Lambda}$ obtained with the covariant chiral potential for the cutoff of $\Lambda_F=550$ (700) MeV is larger (smaller) than the Jul94 result, which corresponds to a stronger (weaker) relativistic effect. This is because the larger the scalar potential, the larger the relativistic effect in the RBHF theory~\cite{Shen:2019dls}.

\begin{figure}[htbp]
    \centering
    \includegraphics[width=0.9\linewidth]{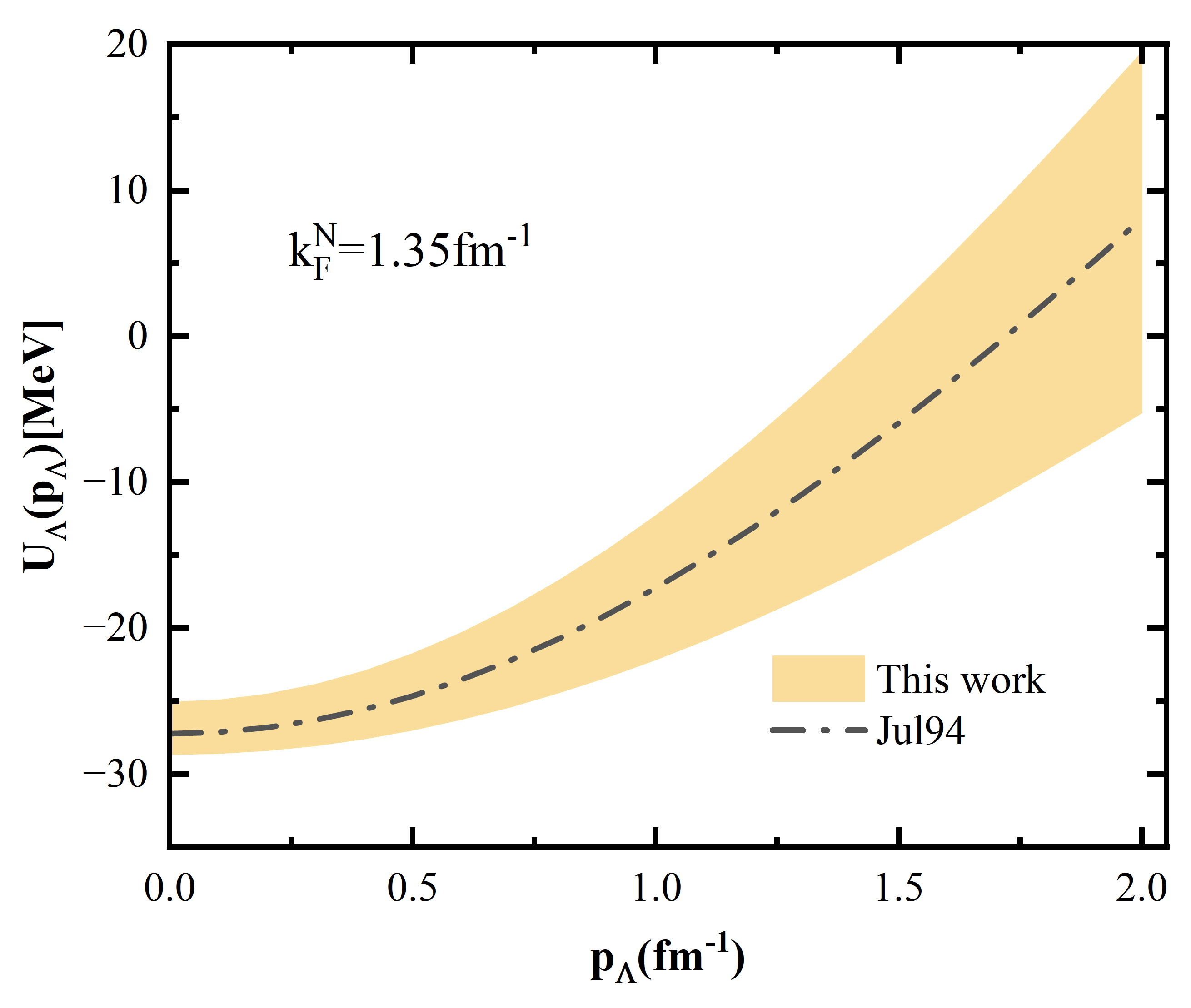}
    \caption{The $\Lambda$ single-particle potential $U_\Lambda(p_\Lambda)$ at $k_F^N=1.35$ $\mathrm{fm}^{-1}$ as a function of the $\Lambda$ momentum, obtained in the RBHF theory with the covariant chiral $YN$ and Jul94 potentials. The band represents the variation of the results with the cutoff, where the lower and upper boundaries represent the results for cutoffs of $\Lambda_F=700$ and 550 MeV, respectively.}
    \label{ulp}
\end{figure}

The momentum dependence of the $\Lambda$ single-particle potential with the nucleon density fixed at the saturation density is presented in Fig.~\ref{ulp}. Our results are also consistent with the Jul94 results within uncertainties. It is seen that these $U_\Lambda(p_\Lambda)$ potentials become more repulsive with increasing momentum $p_\Lambda$. As the $\Lambda$ momentum increases, the uncertainty of the covariant chiral results increases, similar to the behavior of the LO non-relativistic chiral calculations in Ref.~\cite{Petschauer:2015nea}. It is worth mentioning that for the covariant chiral results with the cutoff of $\Lambda_F=550$ MeV (see the upper boundary of the yellow band), the $\Lambda$ single-particle potential becomes positive at relatively low momentum $p_\Lambda \approx 1.4~\mathrm{fm}^{-1}$.

\begin{figure}[htbp]
    \centering
    \includegraphics[width=0.9\linewidth]{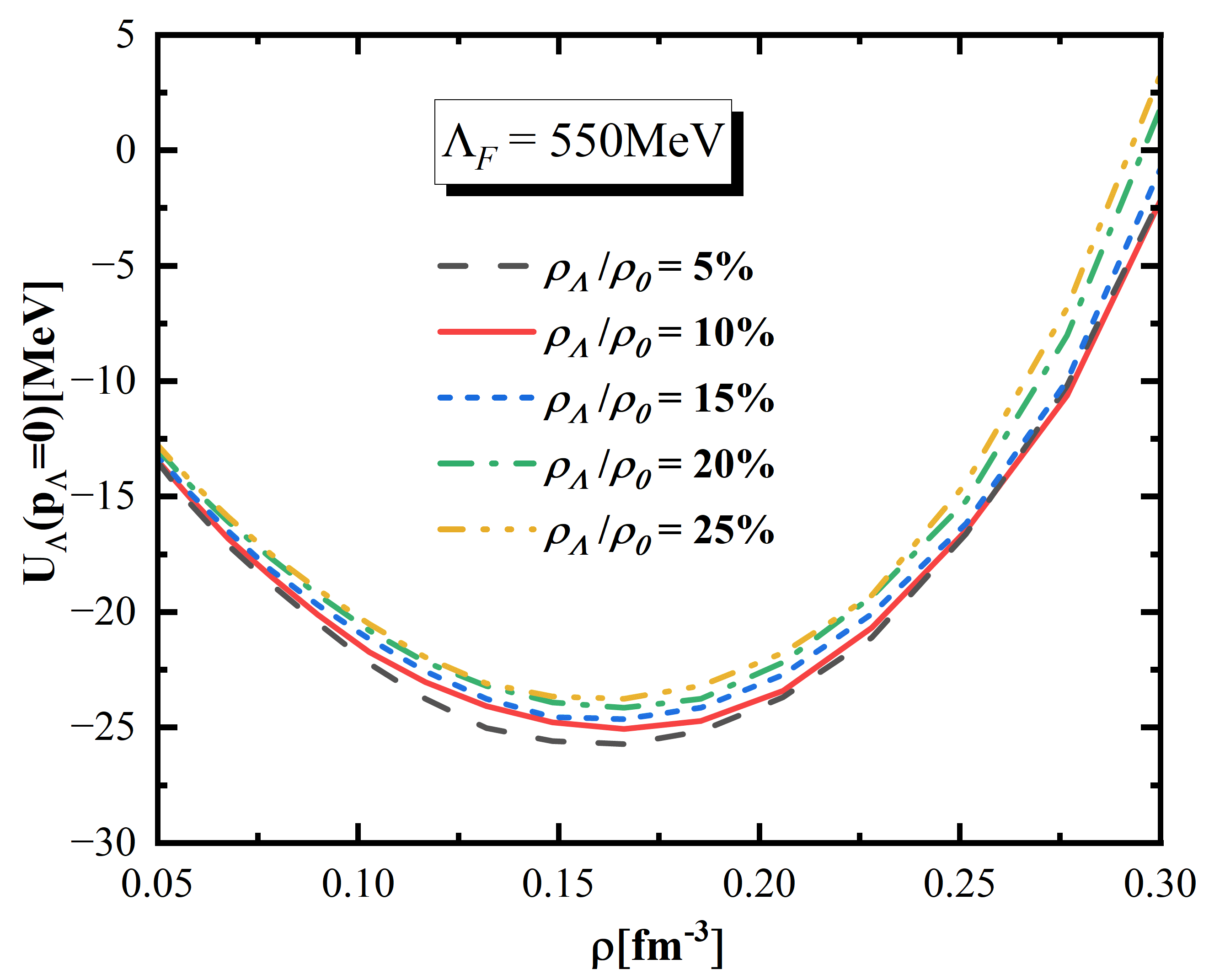}
    \caption{The $\Lambda$ single-particle potential as a function of density in SNM for various $\Lambda$ densities ($\rho_0=0.16~\mathrm{fm}^{-3}$), obtained in the RBHF theory with the covariant chiral $YN$ potential for a cutoff of $\Lambda_F=550$ MeV.}
    \label{kfl}
\end{figure}

It is worth discussing how the $\Lambda$ density affects the calculated $\Lambda$ single-particle potential. In Fig.~\ref{kfl}, we show the $\Lambda$ single-particle potential as a function
of density for different $\Lambda$ densities. These results are calculated in the RBHF theory with the LO covariant ChEFT potential for a cutoff of $\Lambda_F=550$ MeV. The red line represents the $\Lambda$ density used for the results presented in Fig~\ref{ulp=0}. As the $\Lambda$ density increases, the $\Lambda$ single-particle potential becomes more repulsive due to the weaker $\Lambda N$ interaction than the $NN$ interaction. Different $\Lambda$ densities show minor differences at low densities, but the differences become more significant as the density increases. These results are consistent with the Jul94 results of Ref.~\cite{Sammarruca:2009wn}. 

\section{Summary and outlook}\label{Summary and outlook}
In this work, we have studied the in-medium $\Lambda N$ interactions in the relativistic Brueckner-Hartree-Fock theory, employing the leading-order covariant chiral hyperon-nucleon and nucleon-nucleon interactions for the first time. We first re-derived the covariant chiral $YN$ interaction with physical baryon masses and then calculated the $\Lambda$ single-particle potentials $U_\Lambda$ in nuclear matter. On the one hand, a quite satisfactory description of the hyperon-nucleon scattering data points is obtained, comparable with the sophisticated phenomenological models and the non-relativistic NLO ChEFT. On the other hand, the obtained $U_\Lambda(p_\Lambda=0)$ at the nuclear saturation density agrees with the `empirical value.' In addition, we compared our results with those obtained with the non-relativistic ChEFT and phenomenological potentials and found that the results are consistent with the Jul94 results. We also discussed the impact of different $\Lambda$ densities on $U_\Lambda(p_\Lambda = 0)$.

The $\Lambda$ single-particle potential, derived from the relativistic \textit{ab initio} calculation in this work, can serve as a crucial input not only for the hypernuclear structure and neutron star studies but also for constructing the covariant density functional. In addition, we would like to construct the NLO covariant chiral $YN$ forces and investigate their in-medium properties within the RBHF theory. Such work is in progress.

\section*{Declaration of competing interest}
The authors declare that they have no known competing financial interests or personal relationships that could have appeared to influence the work reported in this paper.
\section*{Data availability}
Data will be made available on request.
\section*{Acknowledgements}
We would like to thank Dr. Hui Tong, Dr. Chen-Can Wang, Prof. Xiao-Ying Qu, and Wei-Jiang Zou for many useful discussions. This work is partly supported by the National Key R\&D Program of China under Grant No. 2023YFA1606703 and the National Science Foundation of China under Grant No. 12435007. Zhi-Wei Liu acknowledges support from the National Natural Science Foundation of China under Grant No.12405133, No.12347180, China Postdoctoral Science Foundation under Grant No.2023M740189, and the Postdoctoral Fellowship Program of CPSF under Grant No.GZC20233381.

\appendix



\bibliographystyle{elsarticle-num} 
\bibliography{RBHFY}






\end{document}